# Intersubband polarons in oxides

M. Montes Bajo[a], J. Tamayo-Arriola[a], M. Hugues[b], J. M. Ulloa[a], N. Le Biavan[b], R. Peretti[c], F. H. Julien[d], J. Faist[c], J. M. Chauveau[b], A. Hierro[a,*]

[a]ISOM and Dpto. Ing. Electrónica, Universidad Politécnica de Madrid, Avda. Complutense 30, 28040 Madrid, Spain

[b]Université Côte d'Azur, CNRS, CRHEA, Valbonne, France

[c]Institute for Quantum Electronics, ETH Zürich, Switzerland

[d]Center for Nanoscience and Nanotechnology, Univ. Paris-Sud, University Paris Saclay, Orsay, France

*Corresponding author: adrian.hierro@upm.es (AH)

Intersubband (ISB) polarons result from the interaction of an ISB transition and the longitudinal optical (LO) phonons in a semiconductor quantum well (QW). Their observation requires a very dense two dimensional electron gas (2DEG) in the QW and a polar or highly ionic semiconductor. Here we show that in ZnO/MgZnO QWs the strength of such a coupling can be as high as 1.5 times the LO-phonon frequency due to the very dense 2DEG achieved and the large difference between the static and high-frequency dielectric constants in ZnO. The ISB polaron is observed optically in multiple QW structures with 2DEG densities ranging from $5 \times 10^{12}$ to $5 \times 10^{13}$ cm$^{-2}$, where an unprecedented regime is reached in which the frequency of the upper ISB polaron branch is three times larger than that of the bare ISB transition. This study opens new prospects to the exploitation of oxides in phenomena happening in the ultrastrong coupling regime.

Oxides are semiconductors where a very high control of the crystallinity over the heterointerfaces has recently been achieved allowing to explore phenomena that arise from the presence of a high-mobility two-dimensional electron gas (2DEG) [1,2] as well as from the strong correlation between electrons [3]. Of particular interest has been the ZnO/Mg$_x$Zn$_{1-x}$O material system, where the quantum Hall effect (QHE) has been observed in heterostructures grown by molecular beam epitaxy with electron densities up to $3.7 \times 10^{12}$ cm$^{-2}$ [1]. More recently, the observation of the fractional QHE was also reported in thin layers where the electrons are highly confined and feature mobilities exceeding $1 \times 10^6$ cm$^2$ V$^{-1}$ s$^{-1}$ [3,4], and indeed the rare even-denominator fractional quantum hall states have been identified. [5] Contrary to the QHE, the fractional QHE can only be explained as a result of the presence of a strong correlation between charge carriers driven by a Coulomb interaction. This strong correlation can be highly enhanced if, as in the ZnO/Mg$_x$Zn$_{1-x}$O system, the charge carriers feature a very large effective mass ($0.24m_0$ in ZnO, where $m_0$ is the free electron mass) [6].

The large electron effective mass in the ZnO/Mg$_x$Zn$_{1-x}$O system is further increased as a result of the high ionicity of these polar oxides, which produces a large coupling between electrons and phonons. This strong interaction can be described by a quasiparticle, the polaron [7], with an effective mass of $0.28 - 0.32m_0$ in ZnO [1,8]. Here we consider the intersubband (ISB) polarons, which arise from the collective coupling of the 2DEG with quantized energy levels in the conduction band, i.e. the ISB transitions, and the longitudinal optical (LO) phonons in a semiconductor quantum well (QW). The interaction of ISB transitions with LO phonons has received attention in recent years, both from the theoretical [9,10,11] and the experimental point of view [11,12] [13] [14], with the focus on the GaAs material system. In this paper, we show evidence for ISB polarons at room-temperature in non-polar ZnO/Mg$_x$Zn$_{1-x}$O multiple QWs with 2DEG densities up to 5x10$^{13}$ cm$^{-2}$, homoepitaxially grown on native substrates.

When the interaction between ISB transitions and LO phonons is in the ultrastrong coupling regime, the strength of the coupling is comparable to the bare energy of the excitations involved (in this case the LO-phonon frequency, $\omega_{LO}$), and, in principle, it can enable the observation of a number of fascinating physical phenomena [15,16,17]. Moreover, the coupling between LO phonons and ISB transitions determines the lifetime of carriers in excited subbands [18] and therefore its knowledge is key from the point of view of realizing optoelectronic devices based on ISB transitions, such as quantum cascade lasers [19].

From the microscopic quantum theory of ISB polarons from De Liberato *et al.* [9], the strength of such coupling can be expressed in the semiclassical limit as

$$\Omega = \sqrt{\frac{n_{2D}\omega_{LO}d_{12}^2}{2\varepsilon_0\varepsilon_\rho L_{QW}\hbar}}, \quad (1)$$

where $n_{2D}$ is the 2-dimensional electron density in the QW, $d_{12}$ is the dipole matrix element between the energy levels involved in the ISB transition, $\varepsilon_0$ is the vacuum permittivity, $L_{QW}$ is the thickness of the QW, $\hbar$ is the reduced Planck constant, and $\varepsilon_\rho = 1/(\frac{1}{\varepsilon_\infty} - \frac{1}{\varepsilon(0)})$, with $\varepsilon_\infty$ and $\varepsilon(0)$ being the high-frequency and static dielectric constants, respectively.

Fig. 1 shows $\Omega/\omega_{LO}$ for several semiconductors (see Supplementary Note 1 for details on the calculation), where ZnO stands out with an unprecedently high value signifying that the strong coupling regime should easily be achieved in this material system. In view of Eq. (1), this large $\Omega$ value is a direct consequence of (a) the large difference between the static and high-frequency dielectric constants of ZnO ($\varepsilon(0) = 7.68$ and $\varepsilon_\infty = 3.69$, respectively [20]), which is larger than for most of the usual binary semiconductors (see Supplementary Fig. 1), and (b) the capability of ZnO to accept doping levels as high as $1 \times 10^{21}$ cm$^{-3}$ [21], roughly two orders of magnitude larger than the usual maximum of $\sim 1 \times 10^{19}$ cm$^{-3}$ for GaAs [22].

The ZnO/Mg$_x$Zn$_{1-x}$O material system is also chosen in this work because of the availability of non-polar, native ZnO substrates. Up to now, ISB transitions in wurtzite wide band gap semiconductors, including the GaN and ZnO material systems, have mainly been observed in polar QWs grown along the *c* axis (mostly on foreign substrates, e.g. sapphire or Si) where the quantum confined Stark effect (QCSE) takes place [23,24,25]. Growing the QW structures on a nonpolar orientation (i.e. with the *c*-axis parallel to the interface between layers) can suppress the QCSE [26], and should facilitate the ISB transitions, due to the absence of internal electric fields that have a detrimental impact on the oscillator strength of the transitions. Growing the QW structures on native substrates should also drastically improve the ISB transition-based devices by reducing the defect density [27] and the residual electron concentration of the films, which is as low as $1-5\times 10^{14}$ cm$^{-3}$ in the homoepitaxial m-ZnO/Mg$_x$Zn$_{1-x}$O QWs used in this report [28]. The combination of the extremely high doping capability and low residual electron concentration of ZnO and Mg$_x$Zn$_{1-x}$O provides an unprecedented span of controllable 2D electron concentrations.

**Results and discussion.**

The sample structures presented here are shown schematically in Figs. 2(a) and (b) and summarized in Supplementary Table II. Samples A and B feature 3.7 nm-wide QWs and are Ga-doped $1\times 10^{19}$ and $2\times 10^{19}$ cm$^{-3}$, respectively. Sample C has 4 nm-wide QWs doped $6\times 10^{19}$ cm$^{-3}$ in the central 2 nm of the QW only. An additional undoped sample with nominally identical structure as Sample C is also used as a reference in the reflectance studies. Finally, Sample D features 3.8 nm-wide QWs doped $1\times 10^{20}$ cm$^{-3}$. Self-consistent solving of the Schrödinger and Poisson equations considering also exchange correlation effects [29] yields an energy difference between the first two conduction band confined levels in the QWs, $E_{12}=\hbar\omega_{12}$, very similar in all cases, ranging from 1260 to 1430 cm$^{-1}$ (156 to 177 meV). Fig. 2(c) shows an example of the potential profile of one of the QWs presented here. The frequency of the upper and lower ISB polaron branches calculated for these samples is plotted as a function of the doping in the QW in Fig. 2(d). The upper branch drifts away from $\omega_{12}$ and the lower branch shifts from $\omega_{LO}$ to $\omega_{TO}$ (the frequency of the transverse optical (TO) phonon) as the doping is increased.

The room-temperature 45° IR reflectance spectra of the samples presented in this work are shown in Fig. 3(a) for *p*-polarization of the incident light. In all the spectra, the high-energy side of the reststrahlen band can be seen at ~600 cm$^{-1}$. The bump at ~1500 cm$^{-1}$ is mostly due to the collective interaction between the free electrons along the QW planes (in-plane plasmons) and the light. A reflectance dip is observed, especially in the highly doped samples, that is not present in the spectra taken under *s*-polarization thus fulfilling the polarization selection rule for ISB transitions. As it will be shown below when the results are compared with the model, these features correspond to the upper ISB polaron branch. Its presence in the samples with the lowest 2D electron concentration is also confirmed below with absorption spectroscopy.

Fig. 3(b) shows the reflectance spectra at 45° under *p* polarization in the range where the signature of the lower polaron branch was expected to be found. The main features apparent in all the spectra in the figure are also present in the spectrum corresponding to the undoped reference sample and can therefore be explained by the interaction of the light with the phonon modes of doped ZnO, undoped ZnO, and $Mg_xZn_{1-x}O$ without resorting to coupling with the electrons in the material. The reflectance valley around 525 cm$^{-1}$ is due to the overlap between the reststrahlen bands corresponding to the two phonon modes in $Mg_xZn_{1-x}O$ [30]. The only effect electrons have in these spectra is that of their interaction with the in-plane component of the electric field of the light, resulting in a tilting of the reflectance spectra which, in relative terms, becomes more intense at lower than at larger energies with increasing electron concentration. The reasons why no signature of the lower polaron branch is observable in these experiments are discussed below and require to accurately model the reflectance spectra of all the structures.

The reflectance spectra from all the samples are fitted with transfer matrix models to extract the material-dependent parameters in the dielectric function, which are in turn employed to calculate the dispersion relation for the ISB polaron in these samples. First, the undoped equivalent of Sample C (Figs. 3(c) and (d)) is fitted for both *s* and *p* polarization of the incident light to extract the parameters of the phonon terms of the dielectric function. The shape of the reflectance spectrum of the undoped sample is only due to the interaction of the incident light with the lattice: note that the frequencies of interest for this work lie relatively close to the reststrahlen band, which spans from ~400 to ~600 cm$^{-1}$ in ZnO, i.e. from the $E_1(TO)$ to the $E_1(LO)$ frequencies in the measurement geometry employed here (see Methods section and Supplementary Fig. 3). The phonon-interaction terms thus found are then included in the fits to the doped samples to extract the parameters of the ISB plasmon-light interaction term of the dielectric function (see Figs. 3(c) and (d) for the example of Sample C). Our use of a semiclassical approach with an anisotropic dielectric function to represent the interaction of light with ISB polarons is clearly justified because, as seen in Fig. 3, it reproduces very well the reflectance spectra of the doped sample under both *s* and *p* polarizations, including the frequency and lineshape of the ISB transitions. Supplementary Table II summarizes the obtained values of $\omega_p$ and $n_{2D}$.

The analysis of the reststrahlen band reflectance was repeated for all samples at an incidence angle of 75° and *p* polarization aiming at enhancing the visibility of the lower ISB polaron branch, which should happen when the relative component of the electric field perpendicular to the QW plane is increased. However, all the spectra are again very similar to each other and no sing of the lower ISB polaron branch could be observed (Fig. 4(a)). To find the cause of this absence, Fig. 4(b) shows the reflectance spectra of Sample C together with modelled fits for *p* polarization. If the ISB transition broadening is artificially set to zero, $\gamma_{ISBT} = 0$, it is clear that a reflectance dip appears in the reststrahlen band (i.e. between $\omega_{LO}$ and $\omega_{TO}$). This dip is the signature of the lower ISB polaron. Indeed, when this same simulation is performed for the rest of the samples in the series (Fig. 4(c)) it is found that the reflectance dip shifts to lower energies with increasing 2D electron concentration, consistent with the dispersion relation in Fig. 2(d). However, when the ISB transition broadening of Sample C is reverted to its experimental value, $\gamma_{ISBT} \approx 835$ cm$^{-1}$, the signature of the lower ISB polaron branch in the reflectance spectra becomes undetectable. Note that this is the broadening of the upper ISB polaron branch acting on the lower branch, i.e. a result of the LO phonon-ISB transition coupling itself (Fig. 4(b)). Thus, the lower branch could only be detected in reflectivity in an unrealistic physical system where the ISB transition had zero broadening.

Fig. 5(a) shows the room-temperature absorbance of Sample C in a multipass waveguide configuration for *p*- and *s*-polarizations. In both cases the absorbance increases for low energies due to absorption from free electrons: note in this configuration the light travels along 6 mm of material and, therefore, the samples are essentially opaque for wavenumbers of the light below ~1000 cm$^{-1}$. There is an outstanding feature around 2700 cm$^{-1}$ (335 meV) which only appears in the spectrum measured under *p*-polarization, thus fulfilling the selection rule for ISB transitions. The ISB transiton can be better observed if the ratio between the spectra taken at *p* and *s* polarizations is plotted (Fig. 5(b)). This peak, which corresponds to the upper ISB polaron branch, shifts to higher energies as the 2D electron concentration is increased. Note that the peak energies correspond to what was observed in the reflectance experiments, and also that the ISB transition peak height roughly correlates with the doping level in the QW, consistent with a negligible electron population of the excited level even for the highest doping levels. The full width at half maximum (FWHM) of the measured ISB transitions lies in the range of 800 cm$^{-1}$, i.e. from $\gamma_{ISBT}/\widetilde{\omega}_{12} \approx 0.38$ in the samples with the lowest $n_{2D}$ to $\gamma_{ISBT}/\widetilde{\omega}_{12} \approx 0.28$ in the samples with the largest $n_{2D}$, where $\widetilde{\omega}_{12}$ is the observed ISB transition peak frequency. This is similar to what was obtained by Belmoubarik *et al.* [24] in ISB transition photocurrent experiments at 18 K on c-plane ZnO QWs ($\gamma_{ISBT} \approx 800$ cm$^{-1}$, $\gamma_{ISBT}/\widetilde{\omega}_{12} \approx 0.3$), and lower than the values obtained by Zhao *et al.* [25] in room-temperature absorption experiments, also in c-plane ZnO QWs ($\gamma_{ISBT} \approx 1250 - 1450$ cm$^{-1}$, $\gamma_{ISBT}/\widetilde{\omega}_{12} \approx 0.4$).

The $\omega$ vs $n_{2D}$ dispersion relation of the ISB polaron is now compared to the experimental data, calculated from the absorption spectra of ZnO considering only the off-plane component of the dielectric function (see Methods section). Fig. 6(a) shows the spectral intensity of the ISB polaron modes as a function of the 2D electron concentration in the QWs, whereas Fig. 6(b) shows only the peak wavenumber extracted from the data in Fig. 6(a) for the sake of clarity. It is clear that as $n_{2D}$ is increased, the $\omega_{12}$ and $\omega_{LO}$ resonances drift away from each other as a result of coupling in agreement with Eq. (1). Also, the intensity of the lower branch decreases as the 2D electron concentration is increased, whereas that of the upper branch decreases for low 2D electron concentrations. Both branches have similar intensities in the range of 2D electron concentrations from 10$^{12}$ to 10$^{13}$ cm$^{-2}$, i.e. similar to the $n_{2D}$ ranges presented here. The experimental results for the upper ISB polaron branch extracted from Figs. 3(a) and 5(b) are plotted against $n_{2D}$, which is calculated from Eq. (3) using the $\omega_P$ obtained experimentally from the reflectance spectroscopy. The experimental data agrees excellently with what the dispersion relation predicts. Note the unusual regime reached at these very dense 2DEGs where the observed upper ISB polaron energy is up to three times the energy of the ISB transition itself for the most heavily doped sample. Fig. 6(a) and (b) also include the lower ISB polaron branch energy extracted from the simulated spectra of Fig. 4(c), where the ISB transition broadening was neglected to make the peaks visible. The data point corresponding to sample D is not included since it was not visible in the reflectance spectra of Fig. 4(c) as it falls at the reststrahlen band low-frequency edge and according to Fig. 6(a) its intensity is low at such a high 2D electron concentration.

We are not aware of any other attempt to demonstrate ISB polarons in any other oxide material, although ISB transitions have been already demonstrated in polar ZnO/Mg$_x$Zn$_{1-x}$O [24,25] and, very recently, in MBE-grown SrTiO$_3$/LaAlO$_3$ multiple QWs [31]. There are, however, a number of previous reports on the coupling between phonons and ISB transitions in GaAs-related materials [11,12,13,14], from which the work of Liu *et al.* [12] was probably the first clear and unambiguous demonstration of an ISB polaron. In such work, they demonstrated three-level GaAs/AlGaAs intersubband lasers at a temperature of 80 K where the Stokes Raman shift (i.e. the difference between the pump and lasing energies) was designed to be around the GaAs LO-phonon energy. They found that, instead of having lasing action only when the Stokes Raman shift was equal to the LO-phonon resonance, the device would lase when the Stokes Raman shift was equal to the resonances of a dielectric function analogous to the (off-plane) one employed here. It can be foreseen that the same effect would be observed in a hypothetical ZnO/Mg$_x$Zn$_{1-x}$O three-level lasers designed with the Stokes Raman shift around the LO-phonon energy of ZnO. In the ZnO case, though, the ISB polaronic gap for the Stokes Raman shift would be larger than 14 meV (as opposed to ~6 meV in Liu *et al.* paper) if the transition from the second to the first level of the three-level system is tuned to $\sqrt{\omega_{LO}^2 - \omega_P^2}$. This difference in the width of the ISB polaronic band gap between GaAs and ZnO is mainly due to the larger difference between $\varepsilon_\infty$ and $\varepsilon(0)$ in ZnO.

Finally, the ISB polaron coupling strength is calculated for Sample D (Fig. 1), i.e. the structure with the highest $n_{2D}$, yielding $\Omega/\omega_{LO} \approx 1.5$, unprecedented value that indicates that the strong coupling regime can be reached using the ZnO/Mg$_x$Zn$_{1-x}$O material system.

**Methods**

*Growth of the structures.*

The m-plane multiple QW (MQW) structures presented here were grown by molecular beam epitaxy (MBE) on m-plane ZnO substrates and consist of 10 to 20-period

ZnO/Mg$_{0.3}$Zn$_{0.7}$O MQWs n-doped with Ga at different levels, grown atop a Mg$_{0.3}$Zn$_{0.7}$O buffer layer.

*Transmittance and reflectance spectroscopy*

Slabs 3 mm-wide and 6 mm-long were cut from the as-grown samples, with the short edges oriented parallel to the *c* axis. Multipass waveguides were prepared beveling the short edges at 45° and polishing them mechanically to optical quality with diamond compounds. The back side of the waveguides was polished to optical quality, too. The transmittance was measured at room temperature in a Fourier transform IR (FTIR) spectrometer using a KRS-5, holographic wire-grid polarizer to select the polarization of the incident light. The samples were set in a specially designed blind that allows light only through the beveled edges of the waveguide.

Reflectance measurements were performed at room temperature on the as-grown samples at 45° and 75° angle of incidence of the light in the same FTIR spectrometer. Since the *c* axis lies in-plane for m-ZnO, in *p*-polarization transmittance and in all reflectance measurements we place the *c* axis perpendicular to the electric field (Supplementary Fig. 3), such that the material properties in plane and out of plane are isotropic (except those that arise from the ISB transition) and correspond to the $E \perp c$ orientation.

*Quantum well model*

The values of $\omega_{12}$ in the ISB transition term of the dielectric function are calculated from self-consistent solution of the Schrödinger and Poisson equations for the QWs, including exchange-correlation effects [29]. Both the barrier and QW electron masses are taken to be the ZnO polaron mass $0.28m_0$ [8]. The band offset, $\Delta E_c$, is taken to be $\Delta E_c = 0.675 \cdot E_g$, since the reported values range from 0.65 to 0.70·$E_g$ [24,25,32]. The MgZnO bandgap is calculated, as in Neumann *et al.* [33], using 25 meV per % of Mg at 30% Mg. With this QW model, the oscillator strength of the $e_1 - e_2$ transition, $f_{12}$, is calculated to be between 0.97 and 0.98 in all our samples.

*Reflectance model*

The reflectance spectra are modelled using a semiclassical dielectric function model for each of the constituent layers. For the Mg$_x$Zn$_{1-x}$O layers, the dielectric function is isotropic (i.e. equal for in-plane and off-plane of the QW) and includes only the terms accounting for the interaction between the light and the lattice,

$$\varepsilon_{MgZnO} = \varepsilon_{\infty}^{MgZnO} \prod_{j=1}^{2} \frac{\omega_{LO,j}^2 - \omega^2 - i\gamma_{LO,j}\omega}{\omega_{TO,j}^2 - \omega^2 - i\gamma_{TO,j}\omega}, \quad (2)$$

where $j = 1,2$ stands for the two phonon modes of non-polar Mg$_x$Zn$_{1-x}$O for $\boldsymbol{E} \perp c$, as in Ref. [30], $\varepsilon_{\infty}^{MgZnO}$ is the high-frequency dielectric constant, $\omega_{LO,j}$ and $\omega_{TO,j}$ are the LO and TO phonon frequencies of the $j$ mode, and $\gamma_{LO,j}$ and $\gamma_{TO,j}$ are the LO and TO phonon broadenings of the $j$ mode (see Supplementary Table II for more details on these figures).

In the case of ZnO, the dielectric function is anisotropic, having a different form in the direction parallel to the QW planes, $\varepsilon_{in-plane}$, and perpendicular to the QW planes, $\varepsilon_z$. Both dielectric functions include the background, high-frequency dielectric constant, $\varepsilon_{\infty}^{ZnO}$, and a harmonic oscillator term analogous to that in Eq. (2) to account for the light-phonon interaction (first term in Eqs. (3a) and (3b)). However, $\varepsilon_z$ features an additional harmonic oscillator term to account for the interaction of the light with the ISB transition (second term in (3a)), and $\varepsilon_{in-plane}$ a Drude model term for the in-plane interaction with the electrons (second term in (3b)) [9,34,35]:

$$\varepsilon_z(\omega) = \varepsilon_{\infty}^{ZnO} \frac{\omega_{LO}^2 - \omega^2 - i\omega\gamma_{LO}}{\omega_{TO}^2 - \omega^2 - i\omega\gamma_{TO}} + \frac{f_{12}\omega_P^2}{\omega_{12}^2 - \omega^2 - i\omega\gamma_{ISBT}} \quad (3a)$$

$$\varepsilon_{in-plane}(\omega) = \varepsilon_{\infty}^{ZnO} \frac{\omega_{LO}^2 - \omega^2 - i\omega\gamma_{LO}}{\omega_{TO}^2 - \omega^2 - i\omega\gamma_{TO}} - \frac{\omega_P^2}{\omega^2 + i\omega\gamma_{in-plane}^P}. \quad (3b)$$

$\gamma_{ISBT}$ is a phenomenological ISB transition broadening factor we take to be equal to the FWHM of the experimental ISB transition. $\gamma_{in-plane}^P$ stands for the broadening of the plasma resonance in the in-plane direction, and $\omega_P$ is the plasma frequency defined as

$$\omega_P^2 = \frac{n_{2D}e^2}{m^* L_{eff} \varepsilon_0}, \quad (4)$$

where $\varepsilon_0$ is the permittivity of vacuum, $e$ is the electron charge, $n_{2D}$ is the two dimensional electron density, and $L_{eff}$ is the effective width over which the ISB transition takes place, defined as

$$L_{eff} = \frac{\hbar}{2Sm^*\omega_{12}}, \qquad (5)$$

where

$$S = \int_0^\infty \Psi_1(z)\Psi_2(z) \int_0^z \int_0^{z'} \Psi_1(z'')\Psi_2(z'') dz'' dz' dz, \qquad (6)$$

and $\Psi_1(z)$ and $\Psi_2(z)$ are the wave functions of the first and second confined levels in the QW [36,37]. $L_{eff}$ is $\sim 0.75 - 0.80$ times the QW width in this work (see Supplementary Table II for details).

The reflectance spectra were modelled with the dielectric function approach described above in a transfer matrix formalism (see Ref. 38 and references cited therein) taking into account the whole multilayer structure. The substrate is considered as a semi-infinite medium to account for the light dispersion at the unpolished back-side. The fact that $L_{eff}$ is different from the QW width, $L_{QW}$, is taken into account in the model by setting the central $L_{eff}$ of each QW as doped material and the rest of the QW as undoped material.

The MgZnO in the barrier layers, the undoped ZnO in the substrate, and the fraction of the QW outside $L_{eff}$ are homogeneous and isotropic (i.e. same off-plane and in-plane dielectric function, since $\omega_P = 0$, and Eqs. (3a) and (3b) are then identical). On the contrary, in the doped ZnO in the QW $\omega_P \neq 0$, and it has to be described with both the off-plane (Eq. (3a)) and in-plane (Eq. (3b)) dielectric functions.

Setting $\omega_P = 0$ for ZnO in the QWs of the undoped reference sample, the frequencies and broadenings of the light-phonon interaction terms of both ZnO and Mg$_x$Zn$_{1-x}$O were extracted by fitting the model to the experimental reflectance spectra as shown in Figs. 3(c) and (d), yielding values in agreement with the literature [20,30]. The values thus obtained (see Supplementary Table II) were then used as a reference for the analysis of the doped samples and $\omega_P$ was extracted for all the samples presented here by fitting their reflectance spectra. As discussed in the Supplementary Note 2, the detection limit for the QW 2D electron concentration is approximately $0.3 - 0.6 \times 10^{12}$ cm$^{-2}$, and therefore the technique is not sensitive to the residual electron concentration in these films ($\sim 1 - 5 \times 10^{14}$ cm$^{-3}$).

*Calculation of the dispersion relation from the absorption spectra*

The dispersion relation was calculated from the simulated absorption spectra, $\alpha(\omega)$, as a function of the 2D electron concentration in the QWs. The absorption spectra were in turn calculated from the off-plane dielectric function of the ZnO QWs (Eq. (3a)) as

$$\alpha(\omega) = -Im\left[\frac{\varepsilon_\infty}{\varepsilon_z}\right] n_b \frac{\omega}{c} \frac{\sin^2\theta}{\cos\theta}\,^{38,39}, \qquad (7)$$

where $n_b$ is the background, frequency-independent refractive index, $c$ is the speed of light in vacuum, and $\theta$ is the angle of incidence of the light with respect to the normal to the plane of the QWs.

**Acknowledgements**

We would like to thank Dr. Guzman for his help with the experimental setup and Prof. B. Vinter for the critical reading of the manuscript. This work was funded by the Spanish Ministry of Economy and Competitiveness (MINECO) through Project TEC2014-60173-C2-2, and from the European Union's Horizon 2020 Research and Innovation Program under grant agreement No. 665107 (project ZOTERAC).

**Author contributions**

AH, JTA, MMB, and RP designed the samples. JTA prepared the samples for transmittance and reflectance spectroscopy. MMB and JTA performed the IR spectroscopy. MMB, JMU, and AH performed the numerical simulations. MH and JMC grew the structures. MH, NLB and JMC performed the structural analysis. JMC coordinates the overall project. FHJ and JF participated in the discussions and planning of the study. AH overlooked the whole study. MMB prepared the manuscript and all the authors participated in the discussions leading to this article.

**Competing financial interests**

The authors declare no competing financial interests.

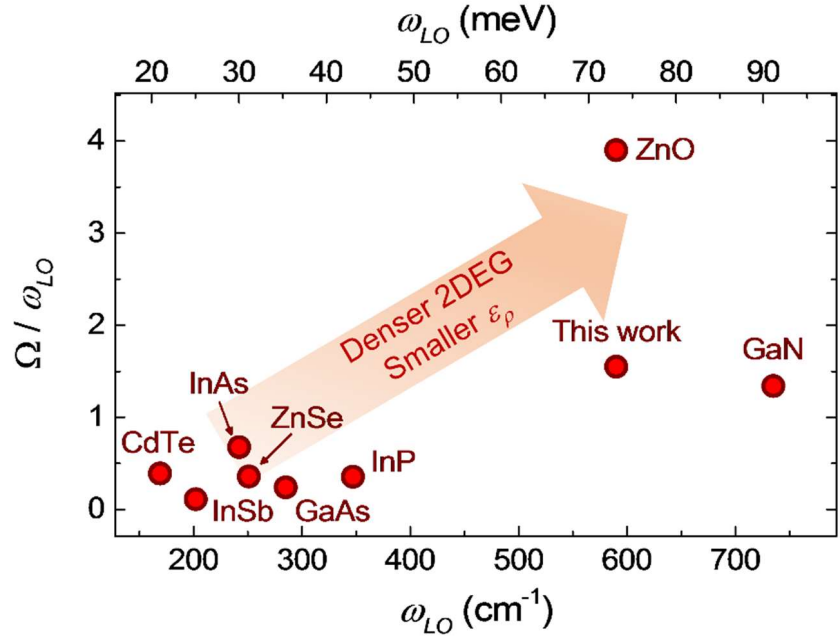

**Figure 1. ISB transition-phonon coupling coefficient for selected semiconductors.** The coupling coefficient $\Omega$, in units of $\omega_{LO}$, is plotted as a function of $\omega_{LO}$ for several binary semiconductors at their respective typical maximum doping. The value of $\Omega$ from this work (sample D) is also included. The figures employed in the calculation can be seen in Supplementary Table I.

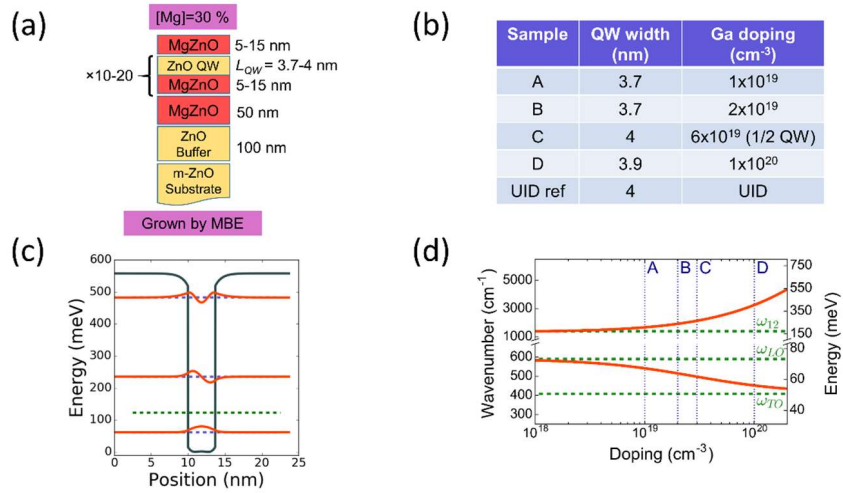

**Figure 2. Details on the samples in this work.** (a) Schematic of the multiple quantum well structures presented in this work. (b) Table showing the different doping levels and schemes employed. (c) Potential profile of one of the QWs in Sample B calculated from self-consistent solving of Schrödinger and Poisson equations considering also exchange correlation effects. The wavefunctions have been offset to match their respective energy levels. The green line indicates the Fermi energy. (d) Dispersion relation for the ISB polarons in these MQW structures as a function of the doping in the QW. The vertical dotted lines indicate the doping of the samples presented here.

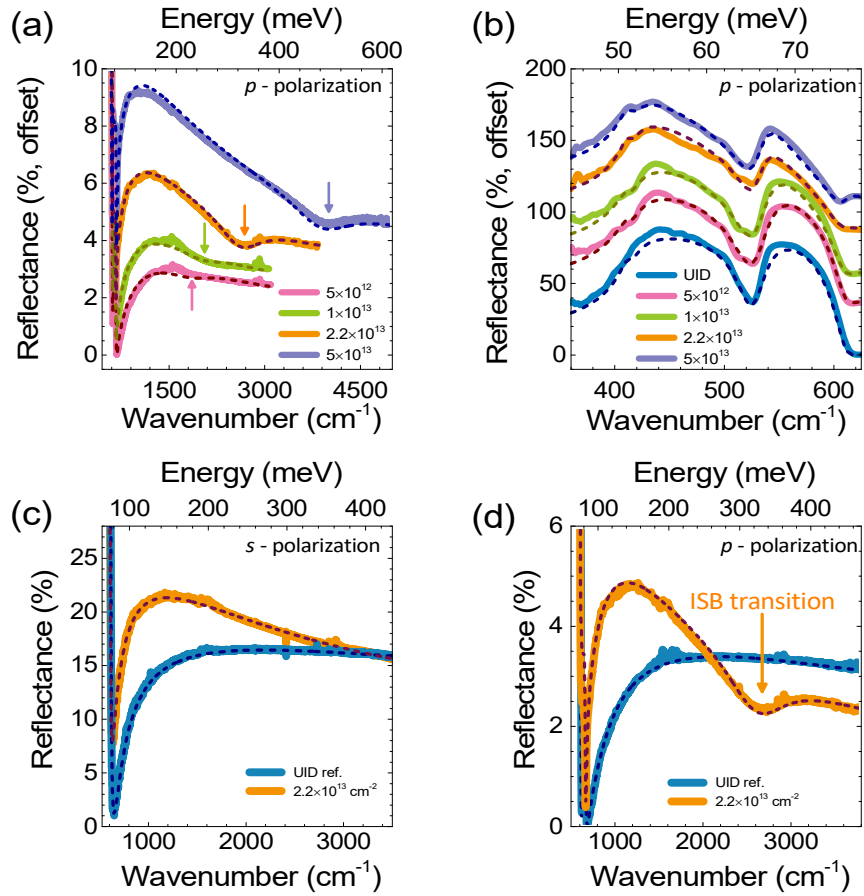

**Figure 3. IR reflectance measurements and fits.** (a) All doped samples in the study under *p*-polarization, offset for clarity. (b) Zoomed-in view of the reststrahlen band in the region where the lower ISB polaron branch should be seen, offset for clarity. (c) Sample D and its equivalent undoped reference under *s*-polarization of the incident light. (d) Same as (c) but under *p*-polarization. The best fit for each spectra is overlaid on the experimental data in all the figures. The legends show the values of $n_{2D}$ extracted from reflectance experiments and Eq. 4 in units of cm$^{-2}$ (Supplementary Table II).

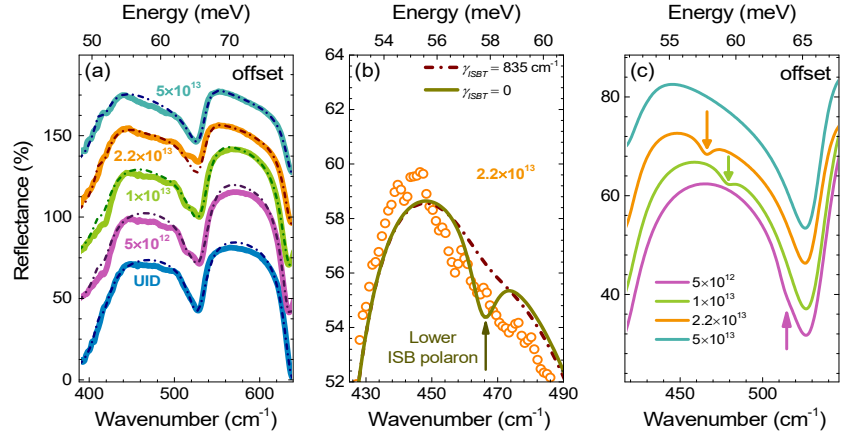

**Figure 4. Visibility of the lower ISB polaron branch on the IR reflectance spectra.** (a) Reflectance of the reststrahlen band at 75°. The spectra are offset for clarity. The experimentally obtained values of $n_{2D}$ are indicated in cm$^{-2}$. (b) Reflectance spectrum of Sample C at 75° and *p* polarization. Overlaid is the best fit of the model with both the actual $\gamma_{ISBT} \approx 835$ cm$^{-1}$ and $\gamma_{ISBT} = 0$. (c) Best fit to the four doped samples in the paper after setting $\gamma_{ISBT} = 0$. Offset for clarity. The energy of the lower ISB polaron branch is indicated with arrows. The legend shows the experimental $n_{2D}$ in cm$^{-2}$ each case. All the curves are obtained under *p* polarization.

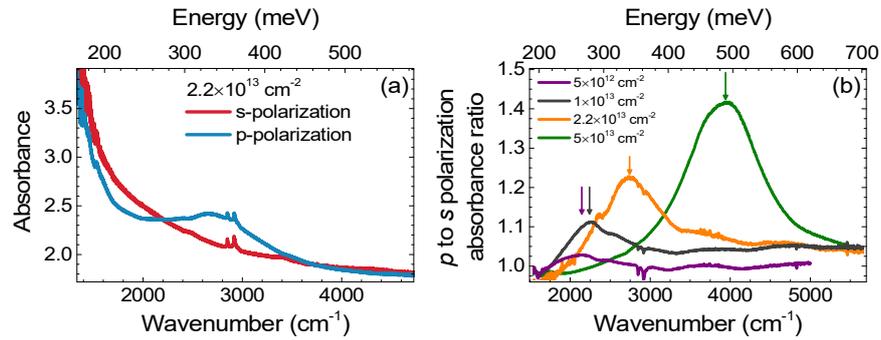

**Figure 5. Absorption transitions of the upper ISB polaron branch.** (a) Absorbance spectra of Sample C under *p*- and *s*-polarization of the incident light in a 45° multipass waveguide configuration. (b) *p*- to *s*- polarization absorbance ratio for all the doped samples in the study. The legends show the experimental values of $n_{2D}$.

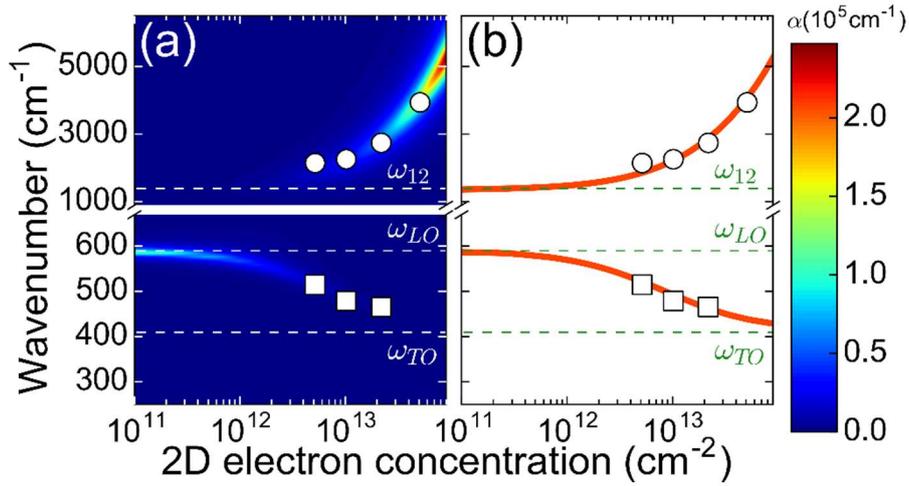

**Figure 6. ISB polaron dispersion relation and match of experimental results.** (a) Dispersion relation for the ISB polaron including the spectral intensity (a) and peak frequency (b) as a function of the 2D electron density in the QWs. The experimental data from the absorbance/reflectance experiments (circles) and the lower ISB polaron energies taken from the calculated reflectance spectra of Fig. 5 (squares) are also shown. The green lines indicate the energy of the LO and TO phonons and the bare ISB transition energy, $\omega_{12}$, i.e. the energy difference between the first confined levels in the QWs.

**Supplementary Information**

**Supplementary Note 1. On the high value of the ISB polaron coupling coefficient in ZnO**

Fig. 1 of the main text compares the value of the ISB polaron coupling strength, $\Omega$, for a number of binary semiconductors. The calculations are made considering the usual maximum value of doping reported for these semiconductors and the maximum reported value for ZnO. Later in the text, $\Omega$ is given for the maximum 2D electron concentration found in this work (i.e. for Sample D). Supplementary Table I summarizes all the values employed in the calculation for each semiconductor.

The high value of $\Omega$ for ZnO can be attributed, on the one hand, to the very large electron concentrations of $1 \times 10^{21}$ cm$^{-1}$ achievable in this materials [1] and, on the other hand, to the very low value of $\varepsilon_\rho$, which in turn is a consequence of the large difference between $\varepsilon(0)$ and $\varepsilon_\infty$. Supplementary Fig. 1 compares these two magnitudes ($\varepsilon_\rho$ and $\varepsilon(0) - \varepsilon_\infty$) for all the semiconductors of Fig. 1 of the main text. ZnO has the largest $\varepsilon(0) - \varepsilon_\infty$ and therefore the smallest $\varepsilon_\rho$ of the whole group.

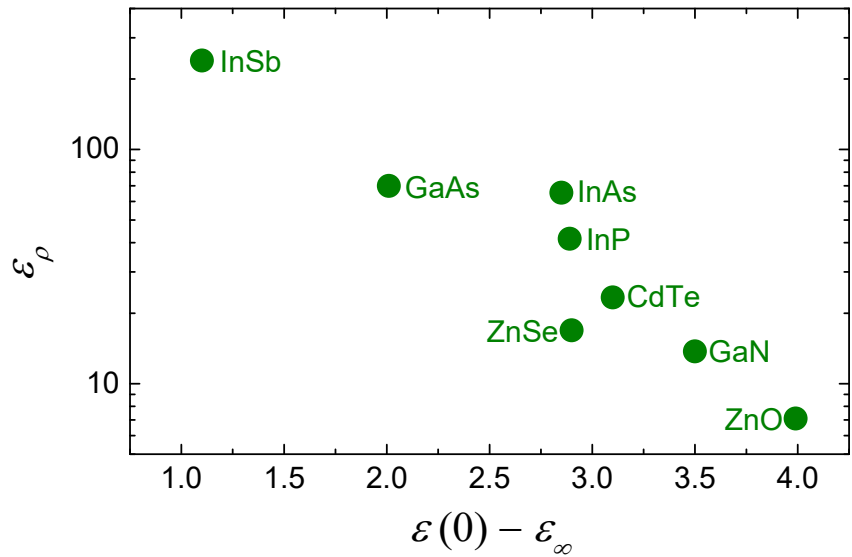

**Supplementary Fig. 1. Comparison of the dielectric constants of several semiconductors.** The reduced dielectric constant, defined as $1/\varepsilon_\rho = 1/\varepsilon_\infty - 1/\varepsilon(0)$, is plotted as a function of the difference between $\varepsilon(0)$ and $\varepsilon_\infty$ for the selection of semiconductors of Fig. 1 in the main text.

|  | Sample D | ZnO | GaN | CdTe | ZnSe | InP | InAs | GaAs | InSb |
|---|---|---|---|---|---|---|---|---|---|
| $n_{3D}$ ($10^{19}$ cm$^{-3}$) | 17 | 100 [1] | 30 [2] | 1 [3] | 0.9 [4] | 3 [5] | 1.2 [6] | 1 [7] | 1 [8] |
| $n_{2D}$ ($10^{13}$ cm$^{-2}$) | 5 | 40 | 12 | 0.4 | 0.4 | 1.2 | 1.8 | 0.4 | 0.4 |
| $\varepsilon_\infty$ | 3.69 | 3.69 | 5.4 | 7.1 | 5.7 | 9.6 | 12.3 | 10.89 | 15.7 |
| $\varepsilon(0)$ | 7.68 | 7.68 | 8.9 | 10.2 | 8.6 | 12.5 | 15.15 | 12.9 | 16.8 |

**Supplementary Table I. Values employed in the calculations of the data in Fig. 1 of the main paper and Supplementary Fig. 1.** The typical maximum achievable values of doping and a QW width of 4 nm were used in all cases except Sample D, where $L_{eff} = 3.0$ nm and the experimentally obtained electron concentration of $n_{2D} = 5 \times 10^{13}$ cm$^{-2}$ were used. Aside from doping values, parameters for all semiconductors were taken from http://www.ioffe.ru/SVA/NSM/Semicond/index.html, except those of ZnO [9], ZnSe [10], and CdTe [11,12].

**Supplementary Note 2. On the detection limit of electron concentration in reflectance.**

The reflectance models shown in this work assume the electron concentration is zero in the undoped layers and, therefore, a value of $\omega_P = 0$ is employed in the expressions for the dielectric function. However homoepitaxial ZnO and MgZnO show a residual electron concentration in the $1 - 5 \times 10^{14}$ cm$^{-3}$ range that raises the question whether this approach is justified and, also, what the lowest measurable electron concentration by the reflectance method employed here would be.

Supplementary Fig. 2(a) shows the modeled reflectance spectra at 45° for a sample with the same structure as Sample C and varying the electron concentration in the ZnO QWs, keeping $\omega_P = 0$ for all the updoped layers. The doping level in the shown spectra spans from $3.8 \times 10^{19}$ to $0.8 \times 10^{18}$ cm$^{-3}$. The spectra are virtually indistinguishable for 2D electron concentrations below $\sim 0.6 \times 10^{12}$ cm$^{-2}$, i.e. doping levels below $\sim 1.5 \times 10^{18}$ cm$^{-3}$, which can be considered the detection limit in our case. Therefore, as the residual electron concentration of undoped ZnO and MgZnO is well below our detection limit, the assumption that $\omega_P = 0$ for undoped layers is correct. Indeed, Supplementary Fig. 2(b) shows how the modelled reflectance spectra for the undoped Sample C considering zero or $5 \times 10^{14}$ cm$^{-3}$ residual electron concentration in ZnO are indistinguishable.

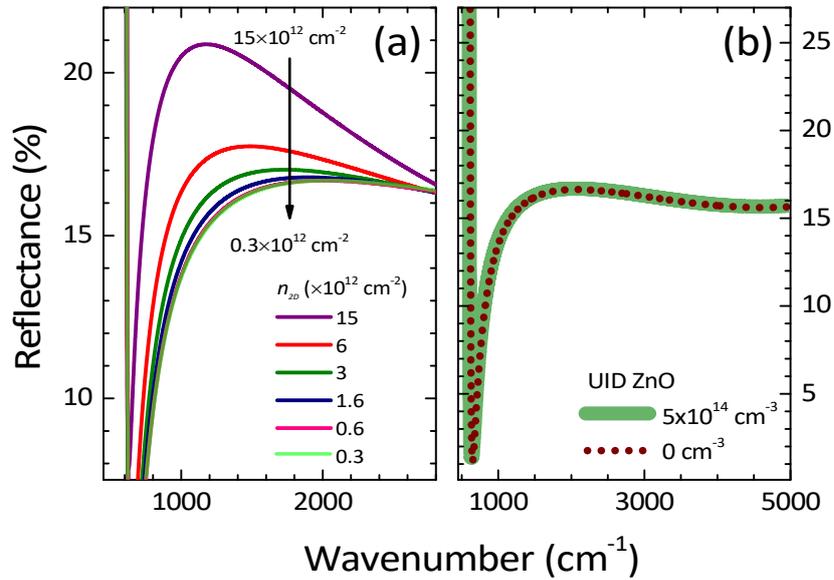

**Supplementary Figure 2. Electron concentration detection limits of the reflectance spectroscopy.** (a) Simulated reflectance spectra of a sample with the same structure as in Sample C and varying electron concentration in the QWs. Note the average 3D doping in the 4 nm-wide QW spans from $3.8 \times 10^{19}$ to $0.8 \times 10^{18}$ cm$^{-3}$. (b) Simulated reflectance spectra of the undoped Sample C assuming zero electron concentration in the ZnO layers (dotted line), or an electron concentration equal to the residual one (solid line).

| Sample | QW width and nominal doping | $\omega_p$ (cm⁻¹) | $\gamma_{ISBT}$ (cm⁻¹) | $L_{eff}$ (nm) | $n_{2D}$ (×10¹³ cm⁻²) |
|---|---|---|---|---|---|
| A | 3.7 nm, $1 \times 10^{19}$ cm⁻³ | 2405 | 820 | 2.9 | 0.5 |
| B | 3.7 nm, $2 \times 10^{19}$ cm⁻³ | 3302 | 843 | 2.9 | 1.0 |
| C | 4 nm, $6 \times 10^{19}$ cm⁻³ (in central ½ QW) | 4269 | 835 | 3.0 | 2.2 |
| D | 3.8 nm, $1 \times 10^{20}$ cm⁻³ | 7205 | 1112 | 3.0 | 5.0 |

**Supplementary Table II. Parameters employed in the dielectric functions for the calculation of the peak absorption coefficient.** An example taken from Sample C of the rest of parameters were: $\gamma_{TO}^{ZnO} = 11$ cm⁻¹, $\omega_{TO}^{ZnO} = 409$ cm⁻¹, $\gamma_{LO}^{ZnO} = 11$ cm⁻¹, $\omega_{LO}^{ZnO} = 590$ cm⁻¹, $\omega_{TO,1}^{MgZnO} = 418$ cm⁻¹, $\gamma_{TO,1}^{ZnMgO} = 17$ cm⁻¹, $\omega_{TO,2}^{MgZnO} = 531$ cm⁻¹, $\gamma_{TO,2}^{MgZnO} = 21$ cm⁻¹, $\omega_{LO,1}^{MgZnO} = 504$ cm⁻¹, $\gamma_{LO,1}^{MgZnO} = 60$ cm⁻¹, $\omega_{LO,2}^{MgZnO} = 663$ cm⁻¹, $\gamma_{LO,2}^{MgZnO} = 18$ cm⁻¹, $\varepsilon_\infty^{ZnO} = 3.69$, $\varepsilon_\infty^{MgZnO} = 3.32$. The table also includes the values of $n_{2D}$ extracted from Eq. 4 in the main text using the values of $\omega_p$ from the reflectance fits.

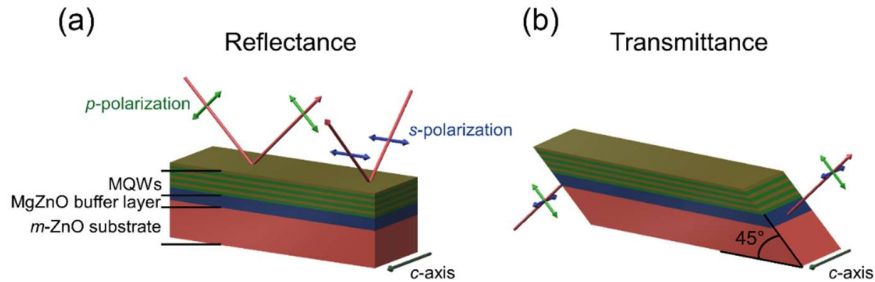

**Supplementary Fig. 3. Measurement geometry for the transmittance and reflectance experiments.** The general structure of the m-ZnO/MgZnO QWs presented here is shown. The incident, reflected, and transmitted light rays are shown together with the respective polarization of the light in each case. Note that in reflectance the electric field is always perpendicular to the *c*-axis.